\documentclass[preprint,12pt]{elsarticle}

\usepackage{amssymb}

\journal{Journal of Magnetism and Magnetic Materials}

\begin{document}

\begin{frontmatter}



\title{Micromagnetic simulations of spinel ferrite particles}


\author{Christine C. Dantas and Adriana M. Gama}
\ead{ccdantas@iae.cta.br; adriana-gama@uol.com.br}
\address{Divis\~ao de Materiais (AMR), Instituto de Aeron\'autica e Espa\c co (IAE), Departamento de Ci\^encia e Tecnologia Aeroespacial (DCTA), Brazil}

\begin{abstract}
This paper presents the results of simulations of the magnetization field {\it ac} response (at $2$ to $12$ GHz) of various submicron ferrite particles (cylindrical dots). The ferrites in the present simulations have the spinel structure, expressed here by M$_{1-n}$Zn$_{n}$Fe$_2$O$_4$ (where M stands for a divalent metal), and the parameters chosen were the following: (a) for $n=0$: M = \{ Fe, Mn, Co, Ni, Mg, Cu \}; (b) for $n=0.1$: M = \{ Fe, Mg \} (mixed ferrites). These runs represent full 3D micromagnetic (one-particle) ferrite simulations.  We find evidences of confined spin waves in all simulations, as well as a complex behavior nearby the main resonance peak in the case of the M = \{ Mg, Cu \} ferrites. A comparison of the $n=0$ and $n=0.1$ cases for fixed M reveals a significant change in the spectra in M = Mg ferrites, but only a minor change in the M = Fe case. An additional larger scale simulation of a $3$ by $3$ particle array was performed using similar conditions of the Fe$_3$O$_4$ (magnetite; $n=0$, M = Fe) one-particle simulation. We find that the main resonance peak of the Fe$_3$O$_4$ one-particle simulation is disfigured in the corresponding 3 by 3 particle simulation, indicating the extent to which dipolar interactions are able to affect the main resonance peak in that magnetic compound.
\end{abstract}


\end{frontmatter}


\clearpage
\section{\label{Intro}Introduction}

Ferrites (ferromagnetic oxides) present convenient dielectric and magnetic properties for microwave and millimeter-wave applications, considering their relatively large magnetic losses and resistivities \cite{Soo60}. It is well known that several physical properties at sub-micron scales, such as the size and shape of the particles in the system, their composition and concentration, including their interactions, are important factors that shape the characteristics of the magnetic material in a sensitive manner \cite{Hye03,Bla06,Mal08}. The main interactions among these  particles are the dipolar (or long range interactions) and the spin exchange interactions. The interplay between these interactions often lead to novel and complex magnetic phenomena. Therefore, in order to design materials appropriate to specific applications, a thorough understanding of these phenomena is needed.

Micromagnetism addresses the study of magnetism at sub-micron scales in the continuum approximation, and its main theoretical equation is the so-called Landau-Lifshitz-Gilbert equation (LLG) \cite{LL1935,Gil1955,Bro1963,Aha96}. It describes the magnetization vector field dynamics (the local precessional motion of the magnetization vector field), including a phenomenological damping term, under an ``effective" magnetic field, representing various interactions amongst the spins. Due to the fact that this is highly nonlinear vector partial differential equation, it is generally solved by numerical methods (analytical solutions can only be found in very few cases \cite{Aha96,Ber01,dAq2004}).

Due to the advance of computer capabilities, micromagetic simulations have been carried out with increasing validity, elucidating several complex magnetic phenomena, but still with many open questions \cite{Aha2001}. In particular, studies of the dynamics of confined spin waves in patterned arrays of magnetic particles in thin films \cite{Jung2002a,Jung2002b,Yu2004,Gub2005,Bai06,Bot09} is of great interest and is the subject of the present investigation, in which ferrite particles are the constituent elements. Although the literature on micromagnetic simulations of ferromagnetic or permalloy particles is quite vast (see, e.g., Ref. \cite{Bai06} and references therein), possibly due to the fact that such a magnetic material is able to support a reasonable range of magnetic structures (specially the vortex structure, relevant to magnetic recording systems), the literature specific on ferrimagnetic or ferrite particle simulations is still somewhat scarce. An inspection of the OOMMF citation list on June 2009 \cite{OOMMF} revealed more than 750 papers that have used that simulator, in which only a few of them focused on ferrimagnetic particles/films (e.g., Refs. \cite{Zie02,Zie04,Zha08}) or bilayers (e.g., Ref. \cite{Zie05}). Zero-field absorption spectra of magnetite cubic particles have been reported in Ref. \cite{Vaa01}. This motivates our project to systematically investigate the magnetization field {\it ac} response of various magnetic compounds (apart from permalloy) according to several physical properties, such as size and shape of the particles, their composition and concentration, inter-particle interactions, etc.

The main purpose of the present work is to study the absorption spectra and magnetization dynamics of full 3D micromagnetic simulations representing submicron  spinel ferrite (e.g., \cite{Soo60}, \cite{Hen07} and references therein) particles (cylindrical dots). Several of these ferrites, already studied for many decades, are now being explored in recent advances in nanotechnology, specially in spintronics (e.g., Ref. \cite{Szo06} and references therein). It is well known that the saturation magnetization of several ferrites can be increased by a proper combination with the non-magnetic zinc ferrite. We have attempted to reproduce qualitatively the effect of an addition of zinc content, in order to have a picture of its possible contributions to the resulting spectrum. In that case, we have focused on a small addition of zinc, which lies in the linear part of the the relation between saturation magnetization and zinc content \cite{Soo60}. This first exploration is intended as a basis for a future systematic numerical work exploring several material properties of ferrites of various types.  

Another relevant analysis in the present work resulted from the performance of an additional larger scale simulation consisting of a $3$ by $3$ particle {\it array}, in a similar fashion to our previous work with permalloy particles \cite{Dan08}. It was performed using analogous conditions of the Fe$_3$O$_4$ one-particle simulation. That larger simulation was performed with the aim of indicating the extent to which dipolar interactions are able to affect the spectrum characteristics in a ferrite patterned film, that is, one formed of closely spaced dots.

This paper is organized as follows. A summary of the time domain micromagnetic simulation setups are given in Sec. II. In Sec. III, we describe the absorption spectra of the simulations and the equilibrium magnetization fields and discuss the results, concluding in Sec. IV.

\section{\label{Simul}Materials and Methods}

\subsection{\label{Theo}The Fundamental Equation of Micromagnetism and Spin Wave Phenomena}

The Landau-Lifshitz-Gilbert equation is a vector partial differential equation for the magnetization vector $\vec{M}$, defined as the sum of $N$ individual magnetic moments $\vec{\mu} _j$ ($j=1, \dots , N$), specified in a elementary volume $dV$ at a position vector $\vec{r}$ within a magnetic system (particle). Being a continuum limit expression, it assumes that the direction of $\vec{M}$ varies continuously with position \cite{Bro1963}. The LLG equation describes the movement of the magnetization field $\vec{M}(\vec{r},t)$ under the action of a external magnetic field ($\vec{H}_{ext}$) as precession movement of $\vec{M}$ around an effective magnetic field ($\vec{H}_{eff}$), defined as $\vec{H}_{eff} \equiv -\mu _0^{-1} \frac{\partial E_{eff}}{\partial \vec{M}}$. It is assumed that $E_{eff}$ embeds the energy of an effective magnetic field, which is in turn generally expressed by the sum of four fields composing spin interactions of distinct origins, namely: $E_{eff} = E_{exch} + E_{anis} + E_{mag} + E_{Zee}$ (respectively: the exchange energy, the anisotropy energy, the magnetostatic or dipolar energy and the Zeeman or external magnetic field energy). The resulting equilibrium state of that system minimizes the total energy. Other physical parameters of the LLG equation are: the saturation magnetization, $M_s$ (determined by the temperature, here fixed throughout), the gyromagnetic ratio, $\gamma$, and a phenomenological (Gilbert) damping constant, $\alpha$. The resulting dynamics is that in which the magnetization vector precesses around the $\vec{H}_{eff}$ field, loosing energy according to the damping term, eventually leading to an alignment of $\vec{M}$ with $\vec{H}_{eff}$. The LLG equation is therefore written as:

\begin{equation}
\frac{d \vec{M}(\vec r,t)}{d t} = 
- \gamma  \vec{M}(\vec r,t) \times \vec{H}_{eff} - 
\frac{\gamma \alpha}{M_s} \vec{M}(\vec r,t) \times 
\left [ \vec{M}(\vec r,t) \times \vec{H}_{eff} \right ].
\label{LLG}
\end{equation}

The magnetization dynamics allows uniform and non-uniform (spatially varying) precession movement within the system. An oscillating magnetic field $H_{ac}$ at a frequency $\omega_0$, applied perpendicularly to the magnetization field leads to a coupling of $\vec{M}$ and $H_{ac}$, in which the energy will be absorbed by the system from the {\it ac} field. The {\it ac} field couples to uniform (leading to main resonance peak) and to nonuniform (spin wave) modes\cite{Pli1999,Neu2006,Bar2007}. In the latter case, one notices that exchange and dipolar interactions may contribute to the energy of these modes. According to the Kittel's model \cite{Kit1958}, additional resonances will be found at frequencies $\omega_p = \omega_0 + D k_p^2$, where $D$ is a function of the exchange  interaction between adjacent spins, and $k_p$ is the (quantized) wave vector corresponding to a given spin wave excitation. It has been noted (e.g., Ref. \cite{Jor00}) that the resonant peaks associated with the exchange interactions are found at the left of the uniform resonance peak, and that dipolar interactions would be generally independent of the size of the system, leading to an interparticle dipolar coupling field (see, e.g., Refs. \cite{Jung2002a,Dan08}). At a reasonably high signal level, they may exist significant coupling between the uniform precession mode and spin wave modes, causing alterations on the main resonance line (e.g., broadening and lowering). A review of confined spin waves can be found in Demokritov et al. \cite{Dem2001}.

\subsection{\label{Setup}Simulations Setup}

The present work follows the general procedure described in our previous work \cite{Dan08}, where a similar analysis of permalloy cylindrical dots has been performed, based in the method outlined by Jung et al. \cite{Jung2002b}. We have used the freely available integrator OOMMF (Object Oriented Micromagnetic Framework)\cite{OOMMF} in order to numerically integrate the LLG equation and evolve the magnetization field of the ferrite particles. The particles were circular dots with finite thickness, that is, cylindrical dots of $0.5 \mu$m of diameter and $85$ nm of thickness. The chosen spinel ferrites for the present simulations (M$_{1-n}$Zn$_{n}$Fe$_2$O$_4$; M, a divalent metal), were the following: (a) for $n=0$: M = \{ Fe, Mn, Co, Ni, Mg, Cu \}; (b) for $n=0.1$: M = \{ Fe, Mg \}. Our simulations were performed in a method suitable for a {\it qualitative} comparison with Brillouin light scattering spectroscopy measurements \cite{Dil82}, noting that formally the strength of the Brillouin cross-section differs quantitatively from the amplitude of the absorption spectrum. We describe the details of the simulations as follows.

In order to study the absorption as a function of frequency, an external magnetic field in the plane of the particle was applied, formed by two components: a static ({\it dc}) magnetic field ($B_{dc} \equiv \mu_0H_{dc}$) of $100$ mT in the $y$ direction, and a varying ({\it ac}) magnetic field ($B_{ac} \equiv \mu_0 H_{ac}$) of small amplitude ($1$ mT) in the $x$ direction:
\begin{equation}
B_{ac} = (1-e^{-\lambda t}) B_{ac,0} \cos (\omega t),
\end{equation}
with the {\it ac} field frequency ($f = \omega / (2 \pi)$) ranging from $2$ to $12$ GHz, in steps of $0.2$ GHz (that is, $51$ different OOMMF frequency runs were performed for each ferrite simulation). We have discretized the time domain of the applied $B_{ac}$ field at intervals of $0.005$ ns, which were used as inputs in the ``field range'' record of OOMMF (stepped linearly by the simulator). The simulations were run up to $5$ ns, resulting in $1000$ outputs (dumps) for each of these $51$ frequency runs. An additional simulation involving a $3$ by $3$ array of particles was performed using the same parameters and conditions of the Fe$_3$O$_4$ one-particle ferrite simulation, but running a smaller set of selected frequencies around the resonance peak, due to the high computational demand of this simulation.

We list in Table \ref{Tab1} the main global parameters of the OOMMF, which were fixed for all sets of simulations. These global parameters were also adopted for the larger scale simulation (the $3 \times 3$ array). Note that the exchange stiffness has a fixed value in all simulations (of the order $\sim 10^{-11}$ J/m). In the next section we address in more detail this parameter in context of exchange length effects. Table \ref{Tab2} lists particular data of the simulations, specifically the value of the saturation magnetization and anisotropy constant adopted for each ferrite particle. We have extracted the data from Fig. 4.9 of Ref. \cite{Soo60}, which presents the experimental values of the saturation magnetization of mixed ferrites (in Bohr magnetons) according to the Zinc content ($n$ value). Data was also extracted from Refs. \cite{Zie05} and \cite{Sel06} (c.f. Table A.1). The simulations were executed on a 3 GHz Intel Pentium PC running Kurumin Linux, taken an average of $\sim 28$ hours of computation for each set of one-particle simulations, whereas the $3 \times 3$ array simulation took about two weeks to be run. 

\subsection{\label{Exchange}Exchange stiffness considerations}

In numerical micromagnetism, it is important to observe the restriction that, in order to obtain accurate results, the value of the computational cell size should not exceed the exchange length (see, e.g. Ref. \cite{dAq2004}), defined as $l_{ex}=\sqrt{2A/(\mu_0 M_s^2)}$, where $A$ is exchange stiffness of the material. Notice the stronger dependence of $l_{ex}$ to $M_s$ than to $A$ (e.g., a $10 \%$ smaller value for $A$ with $M_s$ fixed implies a $\sim 5 \% $ decrease in the resulting $l_{ex}$, whereas the same $10 \%$ reduction in $M_s$ with $A$ fixed implies a $\sim 23 \%$ increase in $l_{ex}$). The cell size here adopted is $5$ nm (enough for a meshing of $\mathcal{O}(100)$ magnetization cells along the particle's diameter), and therefore materials with $l_{ex}$ above that limit are in accordance with the present numerical requirements.

Based on magnetoresistive measurements, Smith et al. \cite{Smi89} obtained the exchange stiffness for Permalloy within $10 \%$ error ($A_{\rm{NiFe}} = 1.05 \times 10^{-11}$ J/m). This method improves on previous estimates based on spin-wave ferromagnetic resonance spectra, which can give discrepancies of a factor of $2$ around the value $A_{\rm{NiFe}} \sim 1 \times 10^{-11}$ J/m. From measurements of the domain width, Livingston \cite{Liv85} found for Fe-Nd-B magnets $A_{\rm{Fe-Nd-B}} = 1.1 \times 10^{-11}$ J/m, but this method depends on the measurement of the anisotropy constant $K1$ (the quote value for $A_{\rm{Fe-Nd-B}}$ was increased by a $\sim 1.5$ factor given a new measurement of $K1$, as mentioned in a note added in proof). Estimates for $A$ can also be obtained by a formula that includes the exchange integral $J$ and other parameters (see, e.g., \cite{Liu07} for an estimate of the exchange stiffness of a nanocrystalline $Ni_{0.5}Zn_{0.5}$ ferrite, although uncertainties are not quoted).

Given the experimental uncertainties, we have decided to adopt an adequate order of magnitude value for $A$ (such that the resulting $l_{ex}$ is above the computational cell of $5$ nm), namely $A \sim 10^{-11}$ J/m. In particular, we have fixed $A= 1.2 \times 10^{-11}$ J/m, as usually quoted for magnetite (Ref. \cite{Zie02}, but see e.g. Ref. \cite{Vaa01} for a quoted value larger by a factor $\sim 3$). Notice that, by the use of a global relation for ferrimagnetic polycrystals \cite{Goy77}, namely, $A(T) = (kT_c/a) (1-T/T_c)^2$, one can alternatively infer the $A(T)$ value for the ferrites from the lattice constant ($a$) and Curie temperature ($T_c$) with $\sim < 20 \% $ precision. Using this relation and data collected from literature (see Refs. \cite{MFT}, \cite{HMAMM}, \cite{Aul65}, and \cite{Zuo06}), we find that the resulting values of $A(T)$ at room temperature for all ferrites in the present work are well within $\sim 20 \%$ of the presently adopted value of $A= 1.2 \times 10^{-11}$ J/m; in other words, reasonably within current experimental uncertainties.

With the adopted value for $A$, we see that $l_{ex} \sim 8.2$ nm for the (M = Fe, n = $0.1$) ferrite (the highest $M_s$ of the set) and $l_{ex} \sim 33.8$ nm for the (M = Cu, n = $0$) ferrite (the lowest $M_s$ of the set). Hence the latter range for $l_{ex}$ is above the cell size, in accordance with the numerical requirements. Notice that, for the most critical cases (namely, M = Fe ferrites, with the highest $M_s$ values), one could ask how much an error in the corresponding $A$ value is allowed for in order to still be in accordance with the numerical requirements (considering that the $M_s$ value is correct). It results that a factor of $\sim 1/3$ (namely, a decrease in $\sim 33 \%$ in $A$) would result in $l_{ex} < \sim 5$ nm for the M = Fe ferrites. We conclude that the adopted value for $A$ is acceptable for the present simulations.

However, it is important to understand how sensitive our simulations are to variations in $A$ to the point that the final results could change appreciably. We will investigate the effect of lower values of $A$\footnote{Clearly, for a fixed $M_s$, lower values of $A$ are of more interest than higher values, since the latter are ``safe" with respect to the numerical requirements for $l_{ex}$, according to our considerations.}  in supplementary simulations to be discussed in the next section along with the main simulations

\subsection{\label{Spec}Calculation of the Spectra}

For the spectra computation, we have also followed the general procedure outlined in Ref. \cite{Jung2002b} (see also \cite{Dan08}). In order to obtain the absorption spectra of the ferrite particles, we proceeded as follows, for each simulation. The first $1$ ns of the averaged magnetization vector in the $x$ direction, $\langle \vec{M} \rangle _{\rm x}(t\leq 1~ {\rm ns})$, has been excluded, and the Fourier transform of the remaining time domain data, $\langle \vec{M} \rangle _{\rm x}(1 < t \leq 5 ~{\rm ns})$, has been calculated. The amplitude of the maximum Fourier peak at each frequency run was then selected for providing the absorpion at the respective frequency, hence building up the spectrum of each ferrite particle.

\section{\label{Resul}Results}

\subsection{Main Micromagnetic Simulations}

In this section we outline the main results found in the present work. 
A more detailed (qualitative) analysis will be offered in the next section.

\begin{figure}
\includegraphics[scale=0.65]{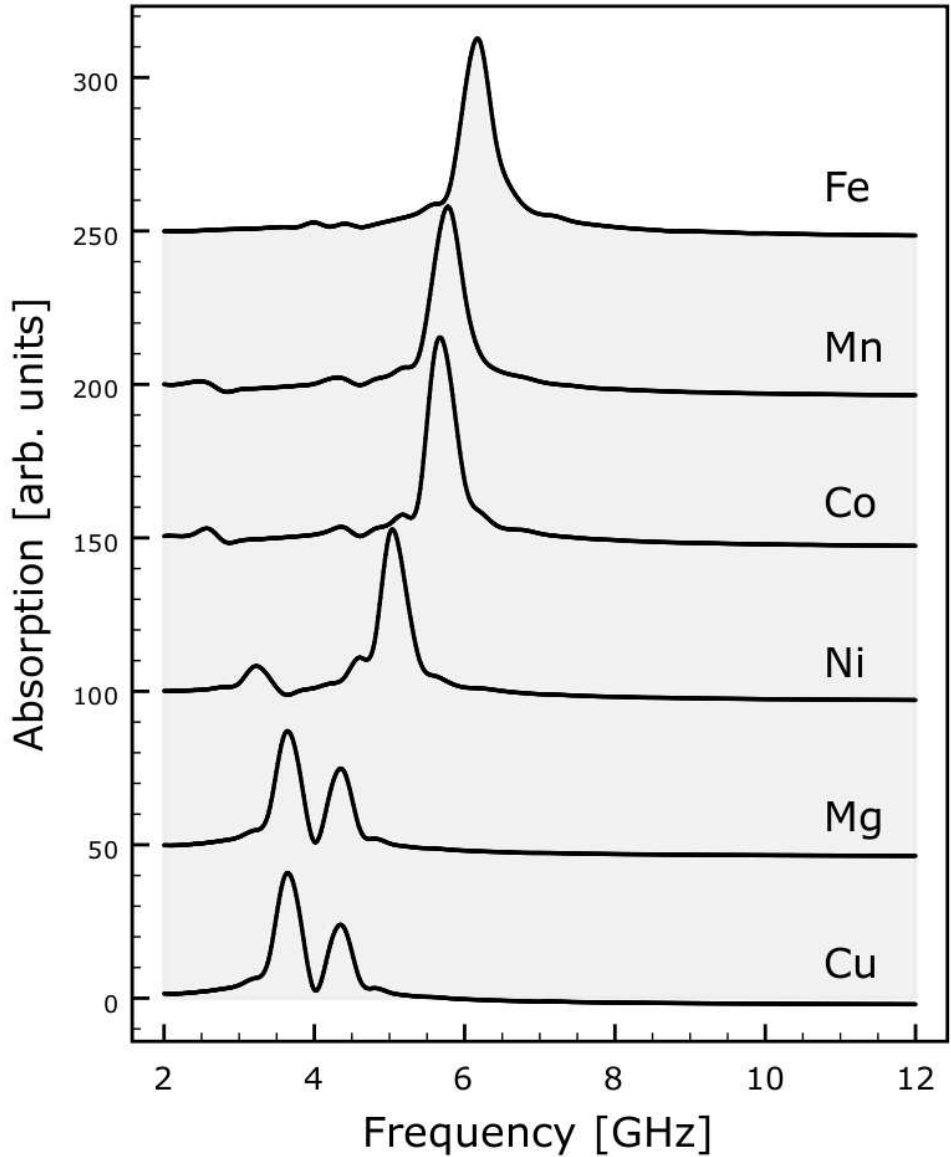}
\caption{\label{fig1} Absorption spectra of single ferrite particle simulations. Curves were arbitrarily dislocated for better comparison. Each ferrite spectrum is labelled by its divalent metal $M$ (all cases here with $n=0$), and organized in order of saturation magnetization, such that the upper curve is from the highest $M_s$. The spectral curves were obtained from spline fits of the discrete simulation results (performed at frequencies from $2$ to $12$ GHz, sampled at intervals of $0.2$ GHz). Possible spin-wave excitations to the left of main resonance peaks can be seen. }
\end{figure}

Fig. \ref{fig1} shows the resulting spline fit absorption spectra of the ferrite one-particle simulations for $n=0$. It is observed that the ferrites with higher saturation magnetization ($M_s$) have their main resonance peaks at higher frequencies. It is also possible to notice in each spectrum the presence of small amplitude absorption peaks at the left of the main resonance peak; these small peaks appear to increase in amplitude for the ferrites with lower $M_s$. It is already pronounced in the M = Ni case, and results in a ``double-like"  peak in the case of the M = \{ Mg, Cu \} ferrites (which have very similar $M_s$). As already mentioned, resonant peaks associated with the exchange interactions are found at the left of the uniform (main) resonance peak (e.g., Ref. \cite{Jor00}).  Hence, the small peaks found in the spectra are probably confined spin-wave excitations of the magnetization field of the particles. 

\begin{figure}
\includegraphics[scale=0.55]{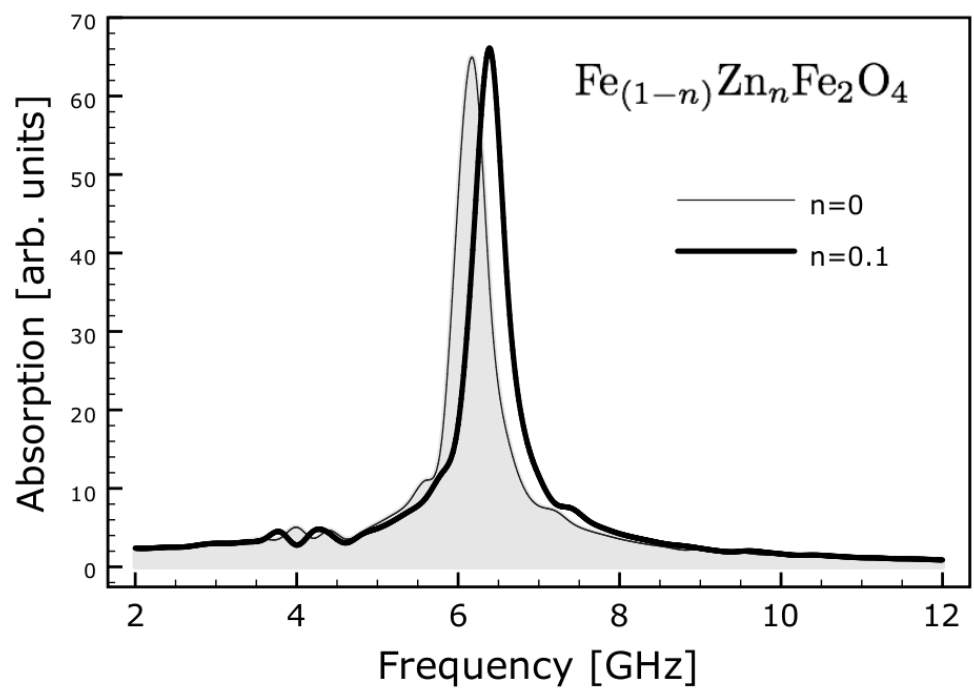}
\caption{\label{fig2} The spectrum of the simulation representing a ${\rm Fe}_{(1-n)}{\rm Zn}_n{\rm Fe}_2{\rm O}_4$ $(n=0.1)$ ferrite compared with that of $n=0$ $({\rm Fe}_3{\rm O}_4)$.}
\end{figure}

\begin{figure}
\includegraphics[scale=0.55]{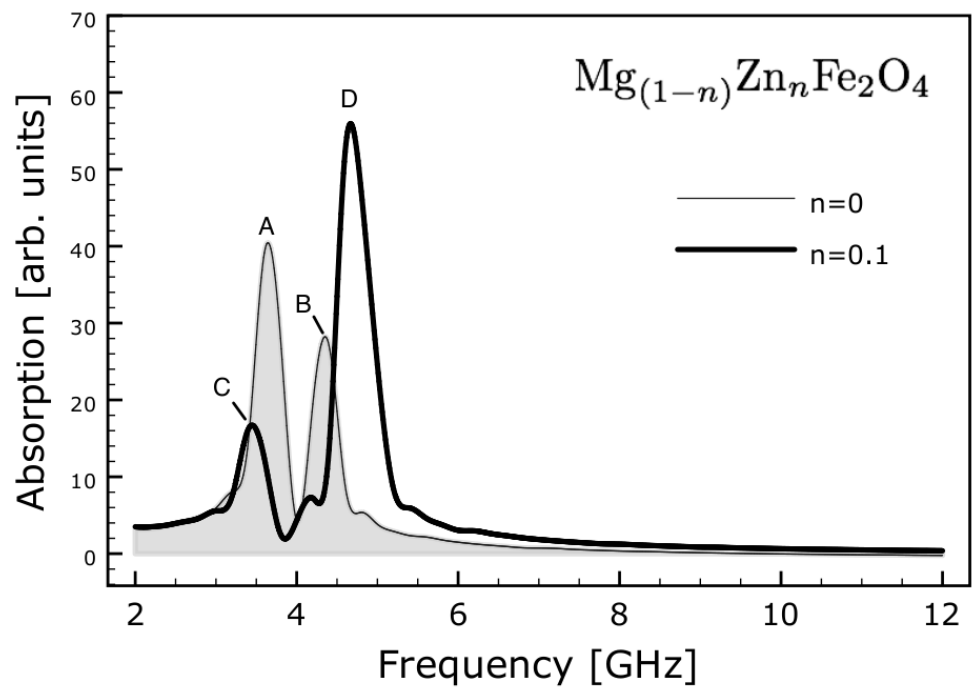}
\caption{\label{fig3} Same of previous figure, but for the ${\rm Mg}_{(1-n)}{\rm Zn}_n{\rm Fe}_2{\rm O}_4$ ferrite. Labels A, B, C and D mark peaks of interest.}
\end{figure}

Fig. \ref{fig2} shows the spectrum of the simulation re\-pre\-sen\-ting a ${\rm Fe}_{(1-n)}{\rm Zn}_n{\rm Fe}_2{\rm O}_4$ $(n=0.1)$ mixed ferrite compared with that of $n=0$ $({\rm Fe}_3{\rm O}_4)$. It can be seen that the main peak in the mixed ferrite is slightly moved to higher frequencies. Fig. \ref{fig3} shows the comparative result for the ${\rm Mg}_{(1-n)}{\rm Zn}_n{\rm Fe}_2{\rm O}_4$ mixed ferrite. This is a more complex case. Clearly, peak A ($n=0$ case), which is quite pronounced, lowers in amplitude significantly in the corresponding $n=0.1$ case (peak C), whereas peak B appears to be modified into peak D, which is at a higher amplitude and frequency. 

\begin{figure}
\includegraphics[scale=0.55]{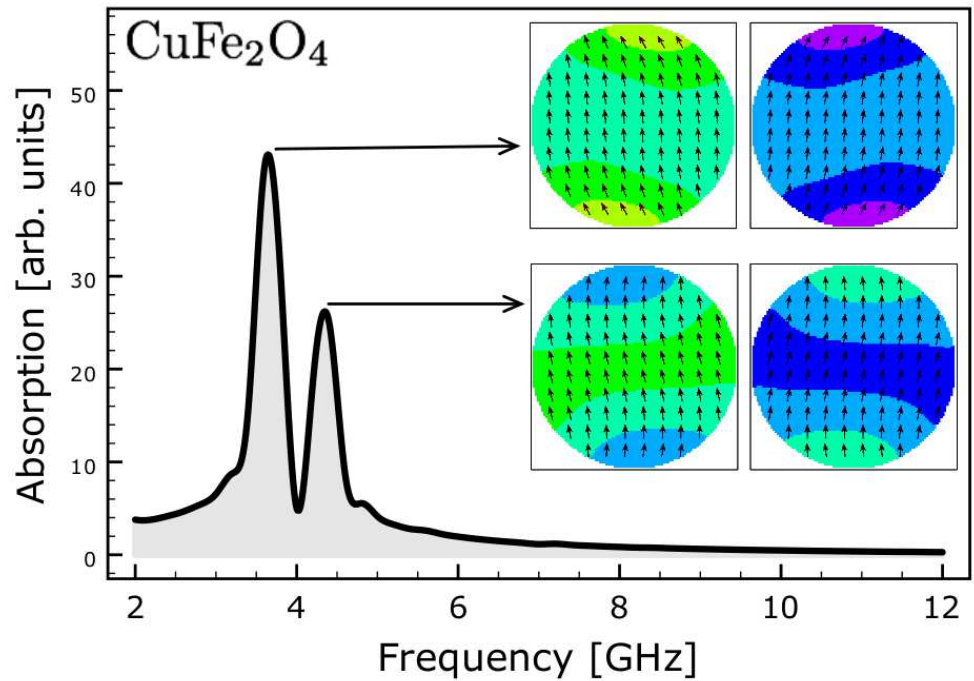}
\caption{\label{fig4} ``Snapshots'' of the magnetization vector field for the ${\rm Cu}{\rm Fe}_2{\rm O}_4$ ferrite, at two resonance peaks, as indicated. At each peak, the snapshots were chosen (restricted to $t > 3$ ns) at two points of the {\it ac} field cycle ($\omega t = \pi/2$ for the snapshot at right, and $\omega t = 3\pi/2$ for the one at left). The varying pixel tonalities of the particle's snapshots correspond to different values of the $x$ component of the magnetization field (subsampled to show an arrow for the average of $9$ vectors per cell element). }
\end{figure}

\begin{figure}
\includegraphics[scale=0.55]{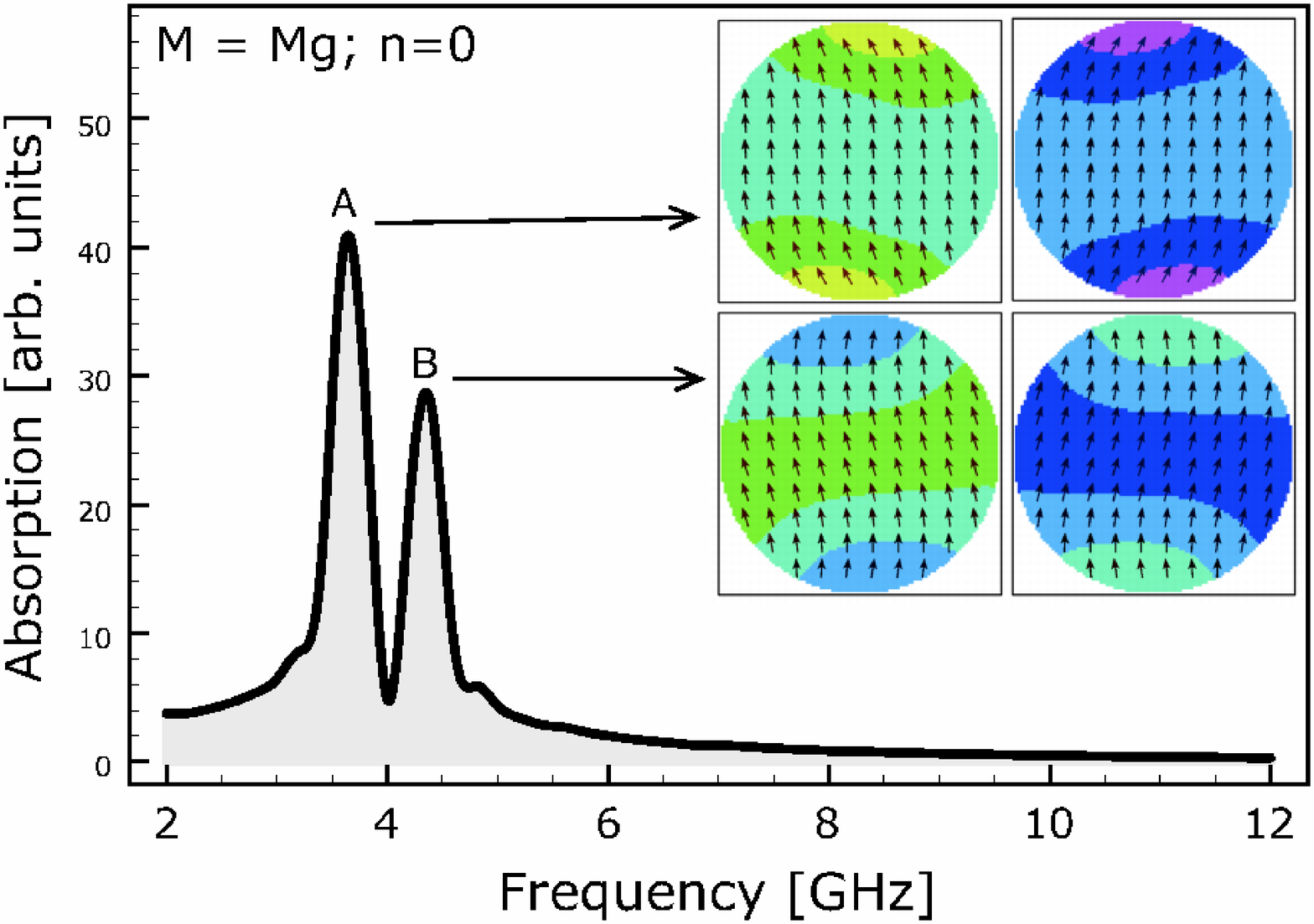}
\includegraphics[scale=0.55]{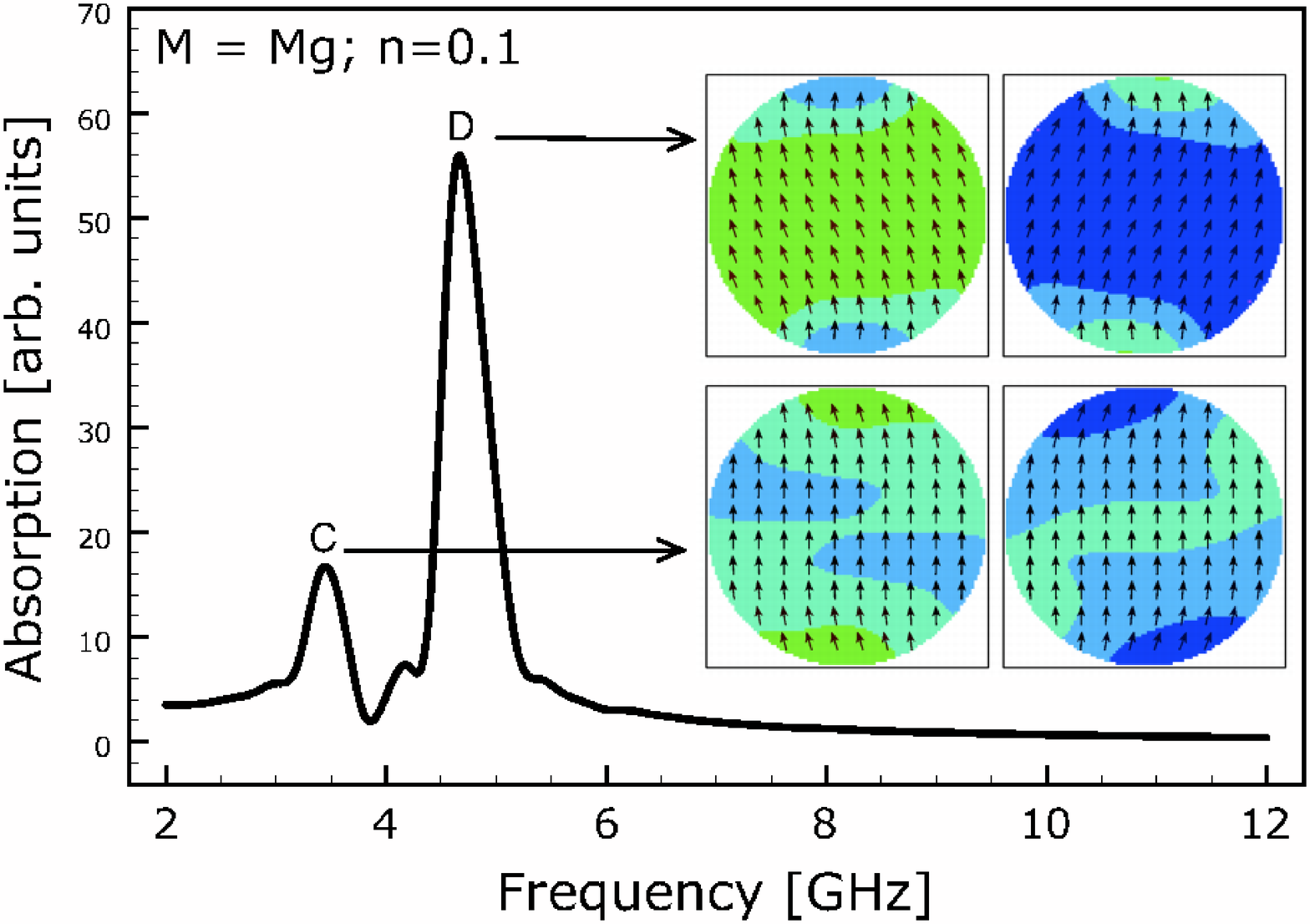}
\caption{\label{fig5} Same as the previous figure, but for the ${\rm Mg}_{(1-n)}{\rm Zn}_n{\rm Fe}_2{\rm O}_4$ ferrite. {\it Top panel:} $n=0$. {\it Bottom panel:} $n=0.1$. }
\end{figure}

In Figs. \ref{fig4} and \ref{fig5} (top panel), the simulation output ``snapshots" of the magnetization vector field related to the peaks of interest of the M = \{Cu, Mg \} ferrites are shown. At each peak, the snapshots were chosen (restricted to $t > 3$ ns) at two points of the {\it ac} field cycle ($\omega t = \pi/2$ for the snapshot at right, and $\omega t = 3\pi/2$ for the one at left of those figures). The varying pixel tonalities of the particle's snapshots correspond to different values of the $x$ component of the magnetization field, which was subsampled to show an arrow for the average of $9$ vectors per cell element. Both simulations show similar results due a close $M_s$ value for these ferrites. It is clear that the pronounced peak at the left of the spectra (in both these cases) is of a different nature from the one at right: in the former peak, the magnetization field in the center of the particle is {\it mostly static and aligned} with the direction of the {\it dc} field; the response of the magnetization field is limited to small oscillations of the (nonuniform) magnetization {\it near the edges}. The corresponding peak at right present instead a {\it quasi-uniform behaviour}.

In both panels of Fig. \ref{fig5}, which refer to a comparison between the $n=0$ and $n=0.1$ cases for fixed M = Mg, one is able to contrast the snapshots of each of the peaks of interest as a function of the zinc content addition. Clearly, the peaks at left (A, C) show a different magnetization field behaviour than the right ones (B, D), as already pointed out in previously. Since the zinc addition is here implemented in a qualitative manner, this result must be interpreted as a general trend.

\begin{figure}
\includegraphics[scale=0.33]{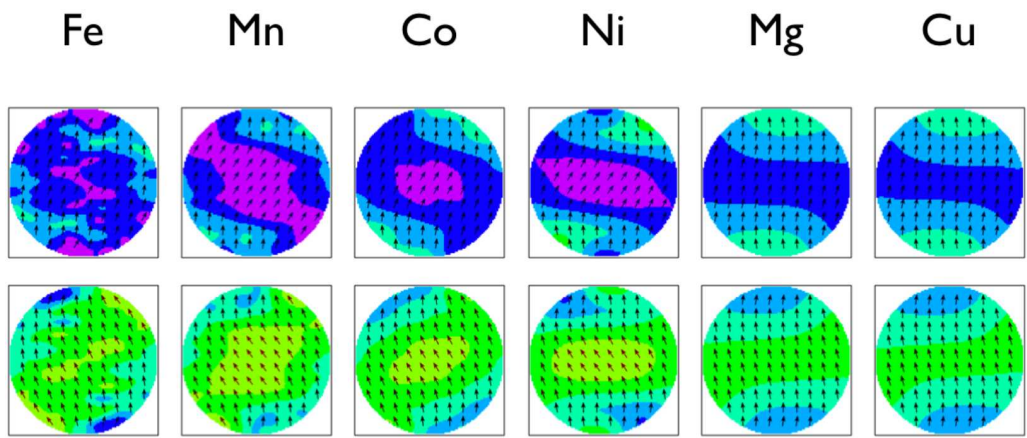}
\caption{\label{fig6} ``Snapshots'' of the magnetization vector field for the ferrites presented in Fig. 1, taken at their main resonance peaks, labelled by its divalent metal $M$ (all cases here with $n=0$). Saturation magnetization increases to the left. In the case of the M = \{ Mg,Cu \} ferrites, the snapshots were selected from the peaks at right of their spectra (see Figs. \ref{fig4} and \ref{fig5}, and the corresponding explanation in the text). As previously, the snapshots were taken at two points of the {\it ac} field cycle ($\omega t = \pi/2$ for upper snapshot, and $\omega t = 3\pi/2$ for the lower one). }
\end{figure}

Fig. \ref{fig6} shows the snapshots of the magnetization vector field at the main resonance peak of each of the one-particle ferrite simulations of Fig. \ref{fig1} ($n=0$). In the case of the M = \{ Mg,Cu \} ferrites, the snapshots were selected from the peaks at the {\it right} of their spectra (see Figs. \ref{fig4} and \ref{fig5} for peaks B and D), given that they show a similar nature with respect to the main resonance peaks of the other ferrites, as already pointed out. It is observed a {\it systematic change in the magnetization field response as a function of the saturation magnetization of the ferrite} (which increases to the left in Fig. \ref{fig6}). This systematic change is revealed in terms of a higher overall amplitude of response for higher $M_s$ as well as an increasingly important presence of small oscillations about a nonuniform static magnetization distribution. Fig. \ref{fig7} is a similar figure to the previous one, but presents instead the snapshots of the simulations with addition of zinc content ($n=0.1$ runs) as compared to their $n=0$ counterparts. We will discuss these results in more detail in the next section.

Fig. \ref{fig8} shows the absorption spectrum of the ${\rm Fe}_3{\rm O}_4$ $3 \times 3$ particles (array) simulation superposed to the corresponding single particle simulation. Clearly, the main resonance peak of the Fe$_3$O$_4$ one-particle simulation is disfigured in the corresponding 3 by 3 particle simulation. There are now 3 resolved peaks approximately within the region of the single main peak of the one-particle simulation counterpart, and these peaks increase in amplitude for higher frequencies, but never reach the same amplitude of the main peak of one-particle run. This figure should provide some indication of the extent to which dipolar interactions are able to affect the main resonance peak in that magnetic compound.
This result is compatible with that of Ref. \cite{Gub2006} for cylindrical Permalloy $3 \times 3$ dot arrays, in which the fundamental mode is found to be split into three modes.

Fig. \ref{fig9} shows the corresponding snapshots of the magnetization vector field at each of the three peaks identified in the previous figure, concerning the 3 by 3 array simulation. Snapshots number 3 (panel at the right of that figure) should be compared with that of Fig. \ref{fig6}, M = Fe one-particle simulation. There are several issues to be observed in the 3 by 3 array simulation, which will be addressed in more detail in the next section.

\subsection{\label{SMS}Supplementary Micromagnetic Simulations}

As mentioned on Sec. \ref{Exchange}, we report on additional simulations performed in order to evaluate the impact of smaller values of the exchange stiffness constant, $A$, on our results. As explained in that section, it is interesting to analyse that impact for the well-known M = Fe ferrite (magnetite). In other words, we have artifically lowered the $A$ value for that one-particle ferrite model by a factor $1/3$ (simulation labeled ``S1") and by $20 \%$ (``S2"), see Table \ref{Sup}. Notice that the S1 run brings $l_{ex} \sim 5$ nm (cell size), and therefore is expected to bring noticeable change in the results (all other parameters remained fixed). Indeed, as Fig. \ref{fig10} shows, there is a decrease in the amplitude of the main peak as $A$ decreases. Otherwise, the resonance frequency and other minor modes at the left of the main peak show little variation. This suggests that, in addition to the preliminary considerations already expressed in Sec. \ref{Exchange}, our results are qualitatively robust.

In order to verify the sensitiveness of the appearance of the ``three peaks" found in the $3 \times 3$ particles simulation with respect to a change in some specific parameter, we have run additional $3 \times 3$ simulations with the same parameters of the original one, except for a change in some parameter of our choice. Due to the fact that these array simulations are computationally demanding, so that a fine-grain covering of the parameter space is prohibitive at this time, we have limited our analysis to a small set of additional simulations in order to infer possible trends. Also, we have limited the simulations to the $5.0$ - $7.0$ GHz frequency range, in steps of $0.2$ GHz. Table \ref{Sup} lists the parameter changed in these simulations (labelled ``S3" to ``S7").

Fig. \ref{fig11} shows the resulting spectra of the additional $3 \times 3$ simulations. The ``Reference" spectrum is that resulted from the original $3 \times 3$ run, that is, the same as shown in Fig. \ref{fig8} (the spectrum with ``three peaks", as indicated). We have also presented a re-analysis of that original simulation by selecting the last $2$ ns of the remaining time domain data, $3 < t \leq 5 ~{\rm ns}$, for the calculation of the Fourier transform  (instead of the $1 < t \leq 5 ~{\rm ns}$ data; see Sec. \ref{Spec}). This selects a clearer steady state condition. We see that (top panel of Fig. \ref{fig11}) the form of the spectrum is practically unchanged, except in amplitude, which is decreased. 

A lower value of $A$ (``S3") also produces a smaller amplitude spectrum, with the overall form maintained (except perhaps for the first, smaller peak at left), which is the same result as the one-particle cases (see Fig. \ref{fig10}). A higher value for $M_s$ (``S4") results in significant distortion of the reference spectrum, namely: a decrease in amplitude of the peaks, and the first, smaller peak at left is not seen in the range of frequencies simulated. An anisotropy constant $K1$ set to zero (``S5"; Fig. \ref{fig11}, bottom panel) shows no significant change in the spectrum. On the other hand, the spectrum resulting from a larger damping constant (factor of $10$) {\it misses entirely the three peaks}. This is interesting in the light of our previous work, where the same damping constant was used, and no splitting of the main peak was found for the Permalloy $3 \times 3$ particles run (see discussions in \cite{Dan08}), although a splitting was indicated in Ref. \cite{Gub2006}. Our present analysis thus confirm that a larger damping parameter possibly explains the difference in the previous results. Finally, the spectrum of the ``S7" run (model B3 of \cite{Dan08}), where the dots ``touch" each other, shows a different spectrum as compared to the reference simulation (which in turn has $0.122 \mu m$ of interparticle spacing). However, a splitting of the main resonance mode is also visible.

\section{Discussion}

The elucidation of peculiar features in the absorption spectra of ferrite particles must take into account recent interpretations on the nature and role of the spin-wave modes. Modes with nodal planes parallel to the magnetization are associated with high frequency modes, whereas modes with nodal planes perpendicular to the magnetization can exhibit frequencies lower and higher than the quasi-uniform mode, and their presence mainly depends on the number of nodes and the equilibrium between the dipolar and exchange interaction effects (see, e.g., Ref. \cite{Jor00}). In the present work, we are interested in the overall qualitative magnetization field {\it ac} response of various ferrite cylindrical dots according to the micromagnetic numerical predictions for these systems, in order to have a basis for a more detailed subsequent investigation. For definiteness, we list here three possible collective {\it ac} responses of the magnetization field according to the following criteria \cite{Jung2002a}:
\begin{itemize}
\item{``Quasi-uniform" behavior ({\it QU}): the motion of each arrow is approximately the same to that of its neighbors, except for the regions around the edge of the particle;}
\item{``Spin-wave" behavior ({\it SW}): the arrows exhibit small oscillations about a nonuniform static magnetization distribution;}
\item{``Edge-like" behavior ({\it ED}): the magnetization field in the center of the particle is mostly static and aligned with the direction of the {\it dc} field; the response of the magnetization field is limited to small oscillations of the (nonuniform) magnetization near the edges -- these modes may be influenced by the dipolar field coming from another particle placed nearby.}
\end{itemize}

The observed characteristics of the absorption spectra as linked to a visual inspection of the magnetization fields at resonant peaks of interest can be classified under those criteria, a subject to which we address now.

A general trend ($n=0$ cases) can be seen by comparing the spectra of Fig. \ref{fig1} with the corresponding snapshots of Fig. \ref{fig6}. The nature of the resonance peaks based on the appearance of the snapshots can be inferred, which we list below:
\begin{itemize}
\item{A remarkable feature is that the main resonance peak in all these simulations seem to be of a similar nature and follow a systematic pattern, namely: an increase in amplitude response (in the central body of the particle) and the presence of small oscillations about a nonuniform static magnetization distribution (SW) -- both effects as a function of a larger saturation magnetization. In the most extreme case, M = Fe, one can see clearly the presence of nonuniformity in the magnetic field oscillation.} 
\item{The anisotropy constant $K_1$ appears to play a minor (but noticeable) role to affect the above-mentioned trend.  For example, let us compare the M = Co ferrite with its two ``neighbours" (in terms of $M_s$ value), namely:  M = Mn and M = Ni (c.f. Fig. \ref{fig6}). If one focuses on the pixel tonalities (a measure of the magnetization amplitude), the M = Co ferrite shows a smaller extent in tonality of the central region (wherein the amplitude of the magnetization field is larger) as compared with those of the M = Mn (which has a close, but larger value of $M_s$) and M = Ni (lower $M_s$) ferrites. It would be expected from the above-mentioned trend (an increase in amplitude magnetization response in the central body of the particle) that such an extent in tonality for the M = Co ferrite would be of intermediate size (between the M = Mn and M = Ni ones). The fact that this is not observed points to a relatively ``easier" alignment of the magnetization with the external field in the M = Co ferrite case. A more systematic study fixing $M_s$ and varying $K_1$, however, was not performed, and more study is needed to confirm these general trends.}
\end{itemize}

\begin{figure}
\includegraphics[scale=0.33]{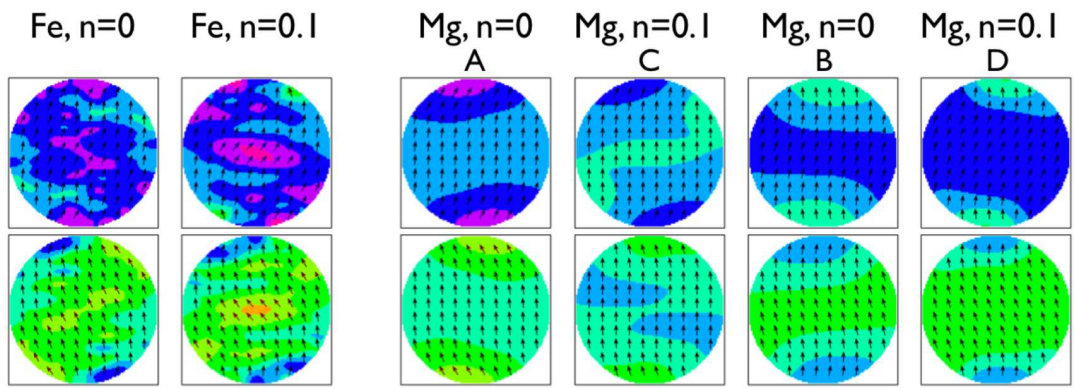}
\caption{\label{fig7} Same of previous figure, but now snapshots refer for the simulations with addition of zinc content ($n=0.1$ runs) as compared to their $n=0$ counterparts. Labels A, B, C and D refer to corresponding peaks in Figs. \ref{fig3} or \ref{fig5}. }
\end{figure}

\begin{figure}
\includegraphics[scale=0.55]{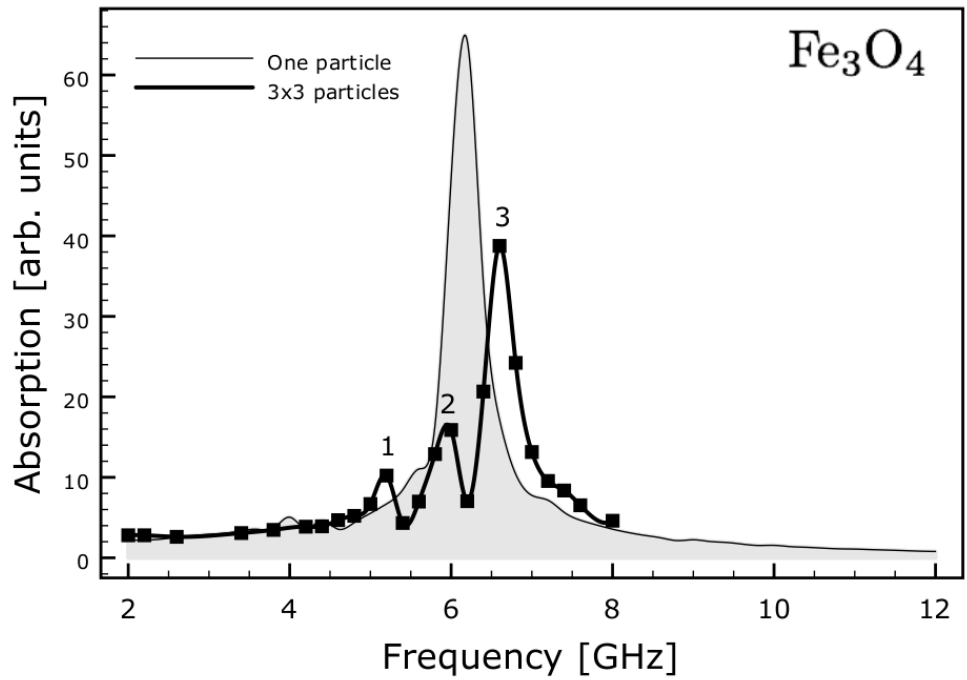}
\caption{\label{fig8} Comparison between the absorption spectrum of the ${\rm Fe}_3{\rm O}_4$ single particle simulation (thin line) with that of the $3 \times 3$ particles (array) simulation (thick line). Filled squares mark the selected frequencies performed for the $3 \times 3$ simulation (the associated curve is a spline fit to the corresponding data).}
\end{figure}

It is specially interesting to observe the snapshots in the cases of the ``double-peaks" seen in the compounds M = \{ Cu, Mg \} (Figs. \ref{fig4} and \ref{fig5}, top panel), where the pronounced peak at right of the spectrum (in both cases) presents an ``edge-like" behaviour ({\it ED}). Such materials will probably show complex spectra in a properly manufactured patterned film, in which``edge-like" modes may be significantly influenced by the dipolar field arising from another particle placed nearby (depending on the interparticle spacing and possibly other factors). We observe that these ``edge-like" effects are {\it smaller for higher saturation magnetizations}. For instance, a comparison between the $n=0$ and $n=0.1$ case (M = Mg; Figs. \ref{fig5} and \ref{fig7} ) clearly shows this effect (note that the $n=0.1$ ferrite has a larger value of $M_s$ than that of $n=0$, for a fixed M). 

A comparison between Figs. \ref{fig2}, \ref{fig5} and \ref{fig7} allow us to qualitatively infer how peaks of interest possibly morph from one to another as the zinc content is added in the ferrite particle. For the case {\it M = Fe}, the nonuniformity ({\it SW}), as expressed by the pixel tonality distribution, tends to {\it increase} in amplitude for $n=0.1$ (or larger $M_s$), specially in the central body of the particle. Similarly, for the {\it M = Mg} case, ``edge-like" effects ({\it ED}) {\it decrease} in amplitude and extent for $n=0.1$ (compare snapshots related to peaks A and C of Fig. \ref{fig7}). For the quasi-uniform modes (peaks B and D of Fig. \ref{fig7}) a {\it more uniform} magnetization field oscillation ({\it QU}) in the central the body of the particle is found for the $n=0.1$ case. These effects qualitatively explain the observed transformation of the corresponding spectra. 

A note is necessary at this point. The $n=0.1$ simulations were intended as preliminar test-cases for a more systematic subsequent work. In Ref. \cite{Gam05}, for example, an experimental study on the overall energy absorption behavior of Mn-Zn mixed ferrites in the frequency range of 8 to 12 GHz for various chemical compositions is presented. A clear understanding of the behaviour shown in that work from the point of view of the magnetic absorption dynamics would be desirable. It would be interesting to analyse and compare the corresponding numerical predictions  with experimental results in order to allow for predictions and guidance for specific applications\footnote{Although the literature on ferrites is extremely vast, it has proven somewhat difficult in the course of this work to find adequate papers to which the present results could be directly compared.}. The present test-cases clearly show that such an analysis is feasible.

\begin{figure}
\includegraphics[scale=0.38]{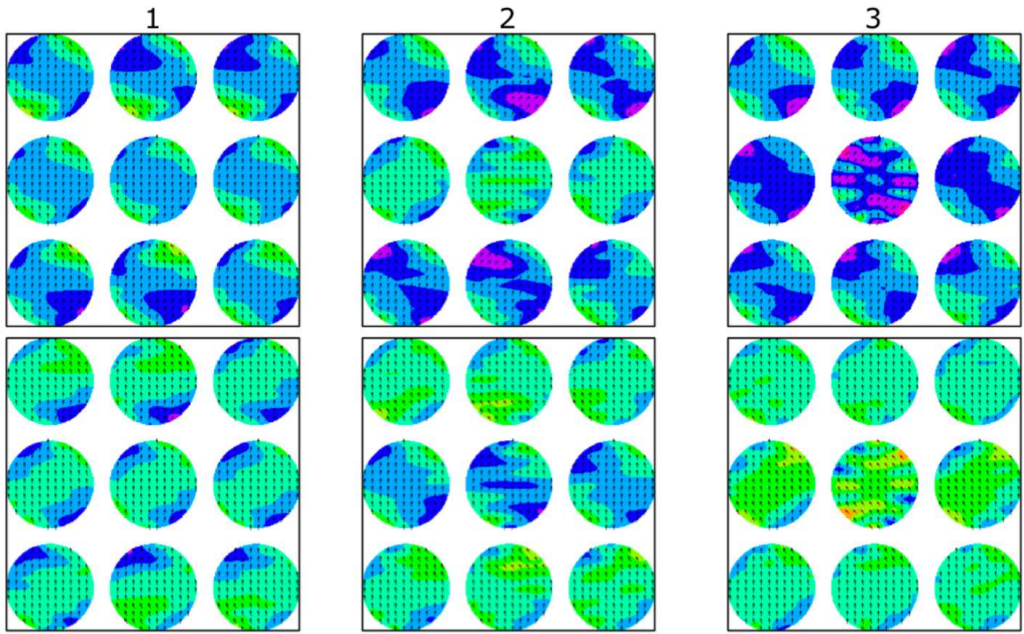}
\caption{\label{fig9} ``Snapshots'' of the magnetization vector field for the $3 \times 3$ particles simulation ($n=0$, M = Fe), obtained at the resonance peaks indicated in the previous figure (labelled by 1, 2 and 3). The snapshots were taken at two points of the {\it ac} field cycle ($\omega t = \pi/2$ for upper snapshot, and $\omega t = 3\pi/2$ for the lower one). Snapshots number 3 (panel at the right) should be compared with that of Fig. \ref{fig6}, M = Fe one-particle simulation. }
\end{figure}

\begin{figure}
\includegraphics[scale=0.55]{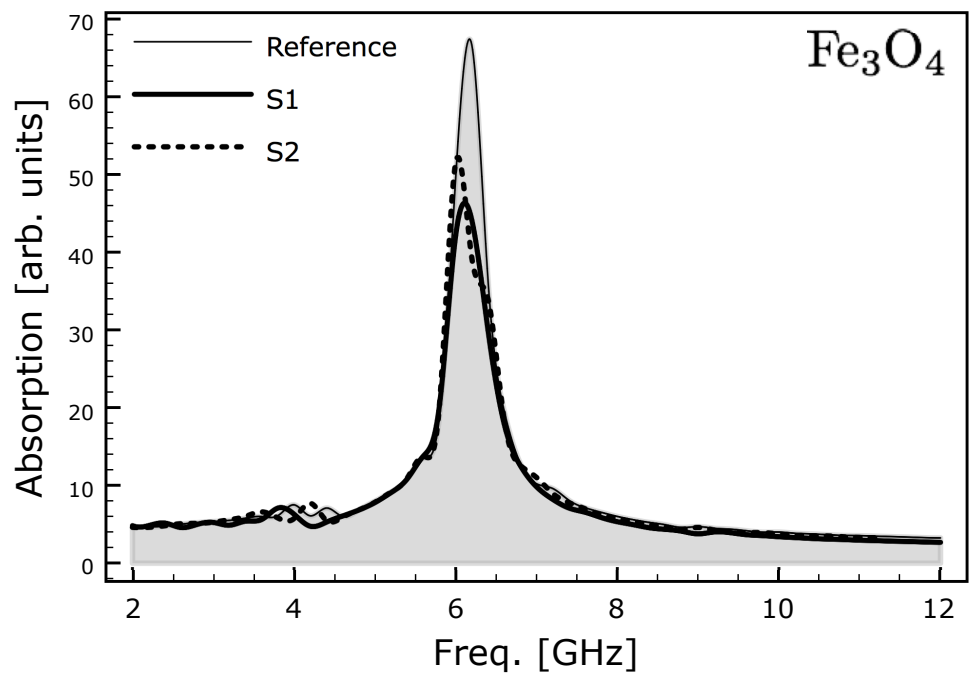}
\caption{\label{fig10} Comparison of the absorption spectra of a single ferrite particle (M = Fe; magnetite), labeled as ``Reference" in the figure, with additional simulations ``S1" and ``S2". The latter runs had their stiffness parameter $A$ artifically lowered the by a factor $1/3$ (``S1") and by $20 \%$ (``S2") in relation to the reference model. }
\end{figure}

\begin{figure}
\includegraphics[scale=0.5]{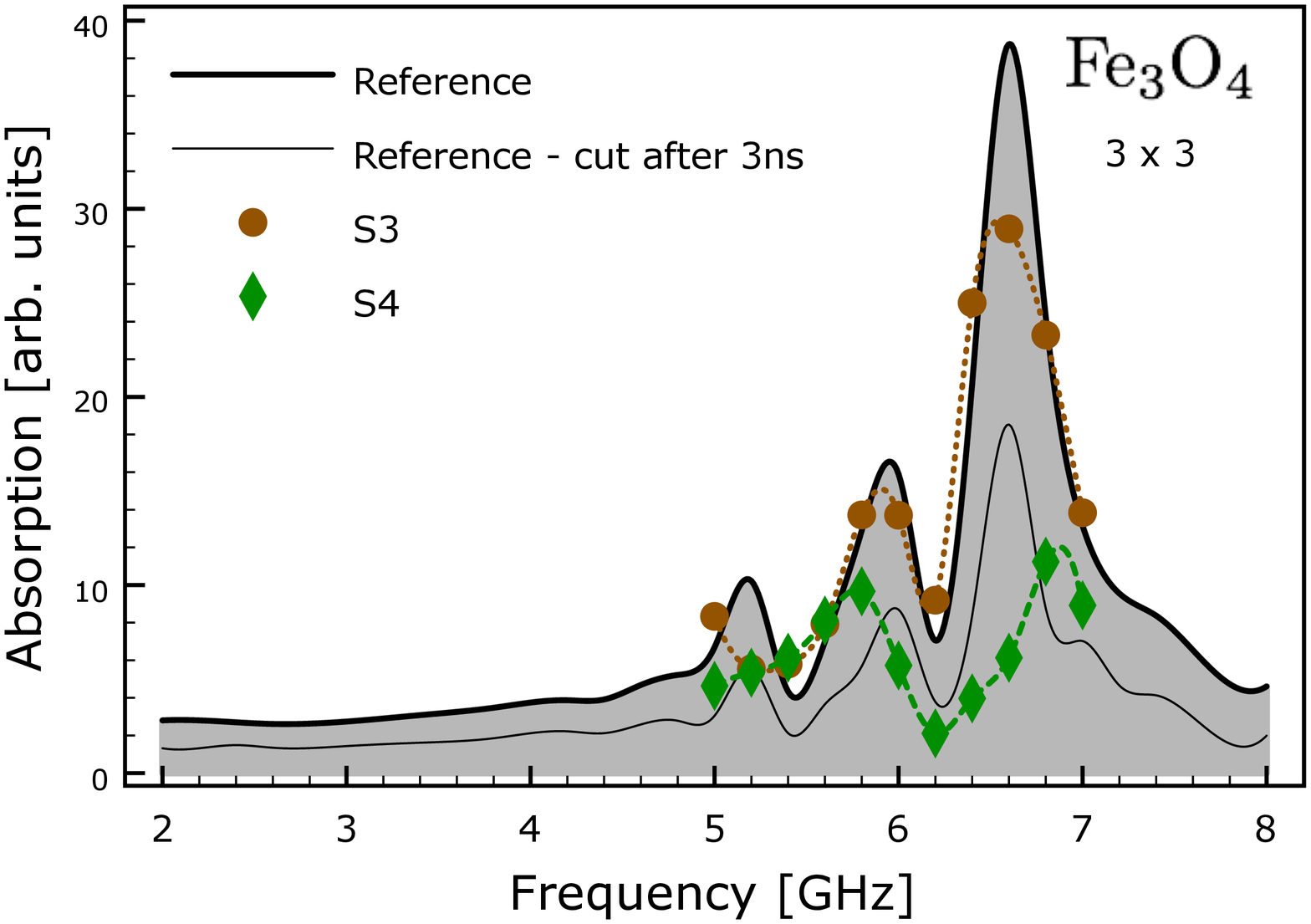}
\includegraphics[scale=0.5]{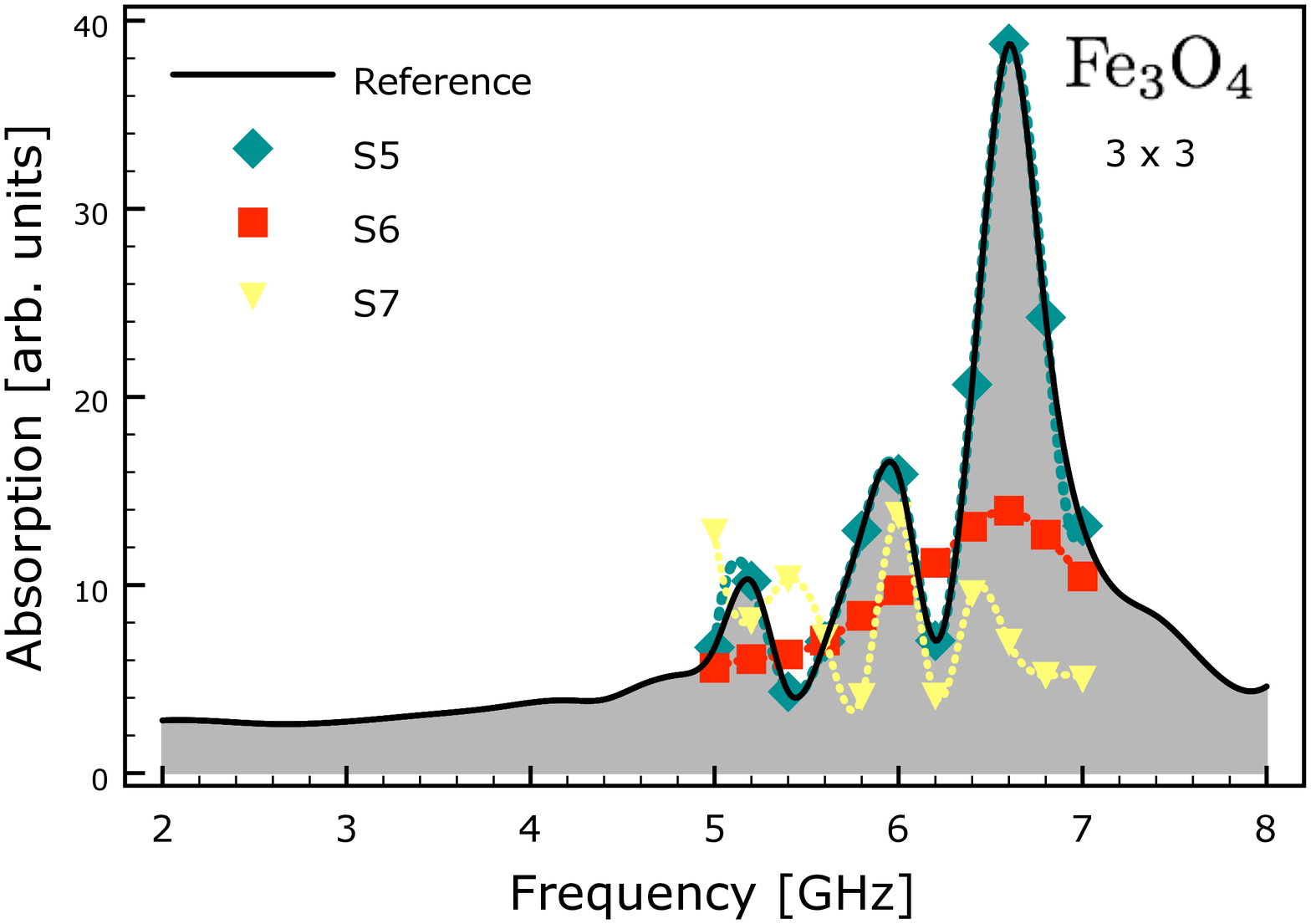}
\caption{\label{fig11} Comparison of the absorption spectra of a $3 \times 3$ ferrite (M = Fe; magnetite) particles simulation, labeled as ``Reference" in the figure, with additional simulations ``S3" to ``S7", as listed in Table \ref{Sup} and explained in the text. }
\end{figure}

The results for the absorption spectrum of the ${\rm Fe}_3{\rm O}_4$ $3 \times 3$ particles (array) simulation is very interesting and follows a previous investigation that we have performed on similar permalloy arrays \cite{Dan08}. As already mentioned, the main resonance peak of the Fe$_3$O$_4$ one-particle simulation is disfigured in the corresponding 3 by 3 particle simulation, resulting in 3 resolved peaks approximately within the region of the original single main peak. Such a feature {\it is not} observed in the permalloy ($M_s = 8.0 \times 10^5$ [A/m]) 3 by 3 array simulation of our previous work (c.f. Fig. 4, the simulation A0 -- one-particle run -- compared with A1 -- 3 by 3 array run -- of that paper, Ref. \cite{Dan08}). A reasonable explanation for not finding the splitting of the main mode in our previous work is due to a larger damping factor used in that work, as suggested here by our supplementary simulations analysis (Sect. \ref{SMS}). A similar splitting was found and discussed in Ref. \cite{Gub2006}. There are, however, similarities between the present and previous results, which we will address now.
\begin{itemize}
\item{The new peaks in the 3 by 3 array simulation never reach the same amplitude of the main peak of one-particle run, and this is also true in the permalloy simulation of our previous work, although only one peak had been observed in that case. As already noted in that paper, the decrease in amplitude of the emerging response is probably due to the averaging out of the magnetization field over the array particles, which show several modes not present in the one-particle case. }
\item{A very interesting effect (c.f. Fig. \ref{fig9}), which we attribute to dipolar interactions, which was found in our previous work and is confirmed here, is the following. Representing the $3$ by $3$ array as a matrix, $\mathcal{A}$, one can observe that the magnetization field of  elements in the $\mathcal{A}_{1,j}$ and $\mathcal{A}_{3,j}$ rows ($j = 1,2,3$) evolve in symmetric opposition to each other. The central row $A_{2,j}$ appears to have its magnetization field evolving independently of the other two.}
\item{The central dot in the 3 by 3 simulation has a particularly unique behaviour compared to the others in the array. Comparing snapshot number 3 (panel at the right of Fig. \ref{fig9}) with that of Fig. \ref{fig6}, which represents the same M = Fe ferrite dot, but  completely isolated, we notice that the presence of small oscillations about a nonuniform static magnetization distribution ({\it SW}), seen in the latter simulation, is {\it only present in that central dot} of the 3 by 3 array. In other words, this effect is apparently attenuated in the other dots of the array, which do gain instead a more ``edge-like" behaviour ({\it ED}).}
\item{Several dots in the snapshots corresponding to peaks 1 and 2 present {\it ED} behaviour, and this effect appears to be more intense in the dots of snapshot number 2 than of number 1, which is consistent with the corresponding amplitude of the peaks in the spectrum (c.f. Fig. \ref{fig8}).}
\end{itemize} 

The effects outlined above must be of a dipolar nature, but the exact prediction of the resulting behavior of the magnetization field (anti-) ``synchronism" as a function of the array symmetry or mutual disposition of the particles still needs elucidation (see Ref. \cite{Dan08} for several examples and discussion).

\section{\label{Conc}Conclusion}

In the present paper we have reported on a set of 3D micromagnetic simulations of cylindrical dots supposed here to represent submicron spinel ferrite particles excited by an external periodic magnetic field. We have analysed the resulting absorption spectra and the magnetization field behavior at modes of interest, limited to the timespan covered in the simulations ($5$ ns). We have identified the nature of confined spin waves and small oscillations of the (nonuniform) magnetization in the absoption spectra through an inspection of the magnetization field at extreme amplitudes of the cycle. A qualitative analysis of the magnetization field behaviour for all simulations was given.

The absorption spectra of ferrite particles may present complex behavior nearby the main resonance peaks, specially in the cases of M = \{ Mg, Cu \} ferrites.  It is inferred that a significant change in the absorption spectrum can be achieved as the zinc content is added in M = Mg ferrites, but this is unlikely in the M = Fe case, at least for a change from $n=0$ to $n=0.1$.  A study of a larger scale simulation of a $3$ by $3$ particle array with similar conditions of the M = Fe  one-particle one shows that the resonance peak of the one-particle ferrite simulation is replaced by a ``triple" peak or otherwise disfigured in agreement with Ref. \cite{Gub2006}. We confirm our previous result that there is indeed a magnetization field (anti-) ``synchronism" effect in the array. This study permitted us to infer the extent which dipolar interactions are able to affect the main resonance peaks in such ferrite particles. 

We aim to perform additional numerical studies to analyse the role of confined spin oscillations in various ferrimagnetic particles and arrays with different physical conditions in a future work.

\section{Acknowledgments}

We thank the referee for useful suggestions. We would also like to thank the attention and technical support of Dr. Michael J. Donahue in the initial phases of this project. We also wish to acknowledge the support of Dr. Mirabel C. Rezende and FINEP/Brazil. 

\clearpage

\begin{table}
\caption{\label{Tab1} Main parameters set to the OOMMF simulator, fixed for all simulations in the present work.}
\begin{tabular}{lll} \hline \hline
Simulation Parameter/Option    & \hspace{1cm} & Parameter Value/Option \\ \hline \hline
Exchange stiffness [J/m]       &  & $1.2 \times 10^{-11}$ \\
Anisotropy Type                &  & cubic \\
First Anisotropy Direction (x,y,z)      &  & (1 1 1)   \\
Second Anisotropy Direction (x,y,z)     &  & (1 0 0)    \\
Damping constant               &  & $0.005$\\
Gyromagnetic ratio [m/(A.s)]   &  & $2.21 \times 10^5$\\
Particle thickness [nm]        &  & $85.0$ \\
Particle diameter [$\mu$m] \footnotemark[1]    &  & $0.5$ \\
Cell size [nm]                 &  & $5.0$ \\
Demagnetization algorithm type &  & magnetization constant in each cell\\
Saturation magnetization [A/m] &  & see Table 2 \\ \hline
\end{tabular}
\end{table}
\footnotetext[1]{There is a small difference in the case of the $3 \times 3$ array simulation. Due to constraints in the drawing of the array (bitmap image to be used as input for the simulator), individual particles turned out to have $0.552 \mu m$ of diameter each, with $0.122 \mu m$ of interparticle spacing, therefore fitting in an exact square of $1.9 \mu m$ by $1.9 \mu m$. It is necessary that the input bitmap size is set as an integer multiple of the cell size.}

\clearpage

\begin{table}
\caption{\label{Tab2} Particular parameter values for each ferrite simulation}
\begin{tabular}{cccc} \hline \hline
Simulation (M,$n$) & $M_s$ [$\times 10^5$ A/m] & K1 [$\times 10^{4}$ J/m$^3$] & Note\\ \hline \hline
(Fe,0.0) & $5.00$ & $-1.10$ & single particle\\
(Fe,0.1) & $5.48$ & $-1.10$ & single particle\\
(Fe,0.0) & $5.00$ & $-1.10$ & 3 by 3 array of particles\\
(Mn,0.0) & $4.14$ & $-0.28$ & single particle\\
(Co,0.0) & $3.98$ & $27.00$ & single particle\\
(Ni,0.0) & $2.70$ & $-0.69$ & single particle\\
(Mg,0.0) & $1.39$ & $-1.50$ & single particle\\
(Mg,0.1) & $2.14$ & $-1.50$ & single particle\\
(Cu,0.0) & $1.35$ & $-0.60$ & single particle\\ \hline
\end{tabular}
\end{table}

\clearpage

\begin{table}
\caption{\label{Sup} Supplementary simulation data}
\begin{tabular}{ccc} \hline \hline
Simulation & Parameter that has been modified & Note\\ \hline \hline
S1         & $A = 0.40 \times 10^{-11}$ [J/m] & single particle\\
S2         & $A = 0.96 \times 10^{-11}$ [J/m] & single particle\\ \hline
S3         & $A = 0.40 \times 10^{-11}$ [J/m] & 3 by 3 array of particles\\ 
S4         & $M_s = 7.5 \times 10^{5}$  [A/m] & 3 by 3 array of particles\\ 
S5         & $K1 = 0$  [J/m$^3$]              & 3 by 3 array of particles\\ 
S6         & damping const. = $0.05$          & 3 by 3 array of particles\\ 
S7         & Model B3 of Ref. \cite{Dan08}    & 3 by 3 array of particles\\ \hline
\end{tabular}
\end{table}

\clearpage

\clearpage

\clearpage

\clearpage

\clearpage

\clearpage

\clearpage

\clearpage

\clearpage

\clearpage





\bibliographystyle{elsarticle-num}
\bibliography{CCDantas_Revised}







\end{document}